\documentclass[conference]{IEEEtran}
\usepackage{graphicx}
\usepackage{graphics}
\usepackage{latexsym,amsmath,epsfig} 
\usepackage{stackengine}
\usepackage{hyperref}
\usepackage{lscape}
\usepackage{amsmath}
\usepackage{amsfonts}
\usepackage{nccmath}
\usepackage{caption}
\usepackage{rotating}
\usepackage{url}

\ifCLASSINFOpdf
\else
\fi
\begin{document}

\title{Simulating Name-like Vectors for Testing Large-scale Entity Resolution}

\author{\IEEEauthorblockN{Samudra Herath}
\IEEEauthorblockA{ARC Centre of Excellence for \\ Mathematical and Statistical Frontiers,\\
University of Adelaide, Australia\\
samudra.herath@adelaide.edu.au}
\and
\IEEEauthorblockN{Matthew Roughan}
\IEEEauthorblockA{ARC Centre of Excellence for \\ Mathematical and Statistical Frontiers, \\
University of Adelaide, Australia\\
 matthew.roughan@adelaide.edu.au}
\and
\IEEEauthorblockN{Gary Glonek}
\IEEEauthorblockA{
University of Adelaide, Australia\\
gary.glonek@adelaide.edu.au}}


%


\maketitle

\begin{abstract}
Accurate and efficient entity resolution (ER) has been a problem in data analysis and data mining projects for decades. In our work, we are interested in developing ER methods to handle big data. Good public datasets are restricted in this area and usually small in size. Simulation is one technique for generating datasets for testing. Existing simulation tools have problems of complexity, scalability and limitations of resampling. We address these problems by introducing a better way of simulating testing data for big data ER. Our proposed simulation model is simple, inexpensive and fast. We focus on avoiding the detail-level simulation of records using a simple vector representation. In this paper, we will discuss how to simulate simple vectors that approximate the properties of names (commonly used as identification keys).
\end{abstract}


%
\IEEEpeerreviewmaketitle

\section{Introduction}

Data integration plays a vital role in data analysis by combining data from different sources into meaningful information. Entity Resolution (ER)  is an important, required step in data integration when identifying a group of records from the same real-world entity in multiple databases. In doing so, records must be matched, but often there is no common key that would make the process easy.

ER is also referred to as the object identity problem, record linkage, the merge/purge problem, de-duplication, duplicate record detection and data matching. 

ER has been widely recognised in academic and statistical research since research data often come from different data sources with different formats. It is of increasing importance in commercial and government practice as well. ER appears as a problem in a range of applications such as medical or epidemiological research, crime detection and national security, tax and other fraud detection, education research, eCommerce applications and customer relationship management \cite{Christen:2012:DMC:2344108}.

A simple illustrative instance of the problem is the process of contact tracing during the novel coronavirus (COVID-19) pandemic. Health experts around the world are currently using contact tracing to determine where patients have caught the virus and whom they might have infected. The process involves tracking patient movements back to the potential person from whom they caught the virus and identifying people who may have been in contact with the infected person. It requires data matching between different databases, especially when only partial information is available on a person. Big data ER techniques will be needed in identifying possible matches and tracing contacts as the number of cases increases. 

The problem dates from at least the 50s, but ER still has open research problems particularly for processing big data. With the increase in very large datasets, the demand for scalable ER techniques has increased. Big data brings new challenges to ER, such as the need for scalability and data quality evaluations. And large-scale test data sets are needed for the development and testing of  ER techniques appropriate for big data. 

Publicly available data sets are usually not big enough or maintained well enough to use as test data in this context. Although there is no strict definition of big data, our interest is in data sets of 10 million -1 billion records. Real datasets are almost impossible to obtain. Apart from the expense, these datasets contain private information. Even if we are allowed to use them for our testing, they can not be showed for replication studies. In the absence of suitable datasets, data simulation is the obvious alternative. 

Ideally, algorithms will ultimately be tested on real data, but there are many advantages in using simulations as part of the development process. A key advantage of simulated data is that the ground truth data, i.e., datasets including the real links, are known. Also, in simulated data, we have control over the properties of the data sets such as size, linking rate, types of errors and error rates.

However, most existing ER simulation tools lack support for the generation of large volumes of data. Our goal is to simulate big datasets suitable for the development and testing of scalable ER techniques. 

We propose a simple, inexpensive and fast simulation model which captures and approximates the most relevant properties of one common identification key (or linking variable), specifically names. The model simulates at the level of detail required to understand the distances between identification keys. We use vectors of numbers to represent records in a dataset aiming these to be low-dimensional. Thus the model allows a simple but flexible definition of the style and parameters of the ER problem and outputs a set of abstract vectors that can be used to evaluate the performance of big data solutions, especially those that involve name matching. 

Our method has several attractive properties when compared to traditional simulation tools. First, to the best of our knowledge, there is no existing simulation tool in the ER literature that can generate millions of records. Traditional simulation tools resample from real databases to generate records. Thus, the size of the existing databases limits the size of realistic samples the model can generate. In contrast, our solution relies on a simple vector representation for records, where the capacity of resampling is effectively unlimited.

Traditional simulation tools are often biased. For instance, existing data identification keys such as names are typically biased towards English, Welsh, Scottish and Irish names and resampling just repeats the bias. We escape this bias using a vector representation that brings universality to our simulation model.

Error modelling is an essential and complex component of test data simulation in ER as errors are inherent in real databases. Creating errors by string simulation require decisions such as types of errors (\textit{e.g.,} spelling errors or phonetic errors), probabilities of introducing errors and the positions of errors. In our simulation model, we generate simple errors using one parameter, which is the variance of errors in the name-like vectors. Indeed, we can introduce small errors with small distances away from original name-like vectors or large errors equally easily.

The main theme of our simulation model is simplicity, which allows us to do inexpensive and fast data generation. In traditional simulation tools, the entire process of error modelling and record generation is often complex, requiring several parameters, user-defined probabilities and other modification to records which increase the computational cost and the complexity of the model. In contrast, our simulation model generates records with $O(d)$ time per record, given the dimension of a vector representation is $d$ (which is small).

The largest dataset of names we found contains 250,000 unique surnames~\cite{7023436}. 
It is natural to question the need for generating 10 million or 100 million names when the namespace is finite and much smaller. Consider, as a motivating example, the applications of ER in linking medical records where one person can have many records over several years. For a country's population, this may be millions or even billions of records and a typical large medical study could potentially involve millions of records.

Our method is one-step towards being able to generate large simulated datasets for big data ER. This will pave the way for us to explore the problem of big data testing of holistic/global level matching. The idea is to avoid detail level pairwise comparisons between records and quickly match the records of big datasets.

We develop our name simulation model based on data analysis of a real namespace. We use three surname datasets for the experimental analysis and the results, closely mirror each other. The simulation model uses a normal distribution to generate name-like vectors in a lower-dimensional Euclidean space. The results indicate that we need a minimum of 6-8 independent random variables to simulate a name-like vector that represents a surname. Also, our model can generate simple edit distance errors by adding noise to the existing Euclidean space using a second normal distribution.

The rest of the paper discusses the methods, experimental analysis and the resulting in contributions toward answering two key questions: (a) ``Can we use vectors to approximate the namespace?'' and (b) ``How do we simulate name-like vectors including errors of a namespace ?'' 

The contributions of this paper: \begin{itemize}
    \item A numerical simulation model to generate name-like vectors based on name dissimilarities including an error~model.
    \item A workflow to evaluate and test the proposed simulation model.
\end{itemize}

\section{Background and Related Work}

The problem of entity resolution using computers started in the early 50s with Newcombe~\cite{Newcombe1959} proposing the automation of ER. Fellegi and Sunter~\cite{Fellegi1969} then described a framework for probabilistic ER in 1969, which is the classic reference in the field. Vatsalan \textit{et al.}~\cite{Vatsalan2013} presented a survey of existing methods and techniques.

A key problem for research in ER is the lack of real-world datasets to experiment with new techniques and to compare existing algorithms. ER, by its nature, requires non-anonymous data but accessing non-anonymous data sets is difficult due to privacy and confidentiality issues. There are frequently used datasets in the ER literature such as a health data set containing midwife data records~\cite{NSW}, the North Carolina Voter database (NCVR)~\cite{NCVR} and CORA \footnote{https://relational.fit.cvut.cz/dataset/CORA}, a citation network, but each of these has limits.

Several simulation tools are described in the ER literature. These tools aim to simulate realistic records and realistic errors that are common in records. Hernandez~\cite{Hernandez1995} presented the first work on data generation creating databases containing duplicate records based on real-world error distributions. Christen~\cite{Christen2005} presented a dataset generator including more attributes and parameters for record generation with more accurate error modelling. This generator is included as a module in the record linkage system called FEBRL (Freely Extensible Biomedical Record Linkage)~\cite{Christen2008}. More simulation studies are presented by Tromp \textit{et al.}~\cite{Tromp2011} and Bachteler~\textit{et al.}~\cite{Bachteler2012TDGen:Methos}.

Most of these simulation methods contain two essential components. Firstly the data are simulated by repeatedly sampling from an existing data set. This limits the size and the controllability of the dataset. Secondly, they introduce errors using models of common error types such as typographical, phonetic and OCR (Optical Character Recognition) errors. The differences between these methods lie in the way they simulate data, simulate errors and the structural models used for introducing errors. The frequency and look-up tables used in FEBRL for its data generation have been regularly used by other simulation studies as well. Even though this is one of the broadly used approaches that is publicly available, it is not scalable enough to explore ER algorithm work with big data sets. We have discussed the general limitations of these traditional simulation tools in the introduction.

Our work is inspired by the vector-based representation of words which has a thriving history in applications of information retrieval, computational semantics and natural language processing. Word embedding maps words to a continuous vector space of a much smaller, fixed-size dimension. The main underlying property is that the words that have similar contextual significance will tend to have similar embeddings~\cite{katharina-siencnik-2015-adapting}. Word2vec in particularly, one of the methods that have been used extensively~\cite{Mikolov2013EfficientEO}. However, even though Word2vec is applicable for natural language databases, it is less useful in proper noun databases such as names, where there is no semantic relationship between names~\cite{Mazeika06cleansingdatabases}. 

There are several methods to embed a set of strings in a metric space including Multidimensional scaling, FastMap, Sparsemap and Lipschitz embedding~\cite{Li-chen}. 
Related works on using such embedding to perform ER can be found in~\cite{Li-chen,DES}. Nevertheless, none of these studies addresses data simulation using a multidimensional Euclidean space where the focus is on cheap comparisons in record matching.

\section{Analysis of Surname Data}

We use three sets of surnames to explore the namespace and its properties, \textit{e.g.,} simple relationships such as the distributions of names and the similarities between name strings. It is surprising to find no studies that describe the relationships between names mathematically. 

The first dataset was collected by Ancestry.com~\cite{7023436}. It contains the 250,000 most commonly occurring surnames in Ancestry.com data. It is the largest unique surname repository we have come across in this study.

The second dataset is collected from the frequency files included in the data set generator program in FEBRL~\cite{Christen2008}. 
The frequency look-up files have been created by extracting values from telephone directories based on Australia. The surname frequency table contains 9000 surnames.

The third data is from the US census\footnote{\url{https://www.census.gov/topics/population/genealogy/data/2010_surnames.html}}. It contains all surnames occurring 100 or more times in the 2010 census, which includes 150,000 distinct surnames. 

In general, all data sets exhibit the same characteristics. For instance, they follow a Zipf's distribution~\cite{zip}. Although the datasets contain surnames from different language groups such as English, Asian and Arabic, they are biased towards English, Welsh, Irish and Scottish. The results of the data analysis are based on the first and the biggest dataset because the others closely mirror the first. 

\section{Methodology}

Our aim is to find a simple representation for names so that we can simulate them quickly. We want to use the data for matching, so the main property of interest is the similarities or dissimilarities between names. We assume that names reside on a manifold in a very high-dimensional space (the namespace) where similar names are close together while dissimilar names are far apart.  The first key question is: ``Can we use lower-dimensional vectors to approximate the namespace?" In other words, we seek to approximate a high-dimensional manifold with a lower-dimensional Euclidean space.


We cast the problem as a dimensionality reduction problem, and employ least-squares multidimensional scaling (LSMDS)~\cite{Kruskal1964} to address the problem. By using LSMDS, we can project names into a lower-dimensional Euclidean space approximately preserving the dissimilarities between them. Then the pairwise comparisons between names become Euclidean distances between vectors. This lower-dimensional representation of namespace allows us to study the structure and explain the relationships mathematically.

\subsection{Multidimensional Scaling}


Multidimensional Scaling (MDS) is a group of approaches to map a set of points described by a dissimilarity matrix into a Euclidean space. In MDS, a configuration is defined as $n$ data points in $p$ dimensions. Among the variants of MDS, we use LSMDS as it worked best with our data.

\subsubsection{Least-squares multidimensional scaling}

The standard method in LSMDS is to minimize the raw stress ($\sigma_{raw}$) of a configuration. The input is a dissimilarity matrix \mbox{$\Delta=[\delta_{ij}]$}, where $i$ and $j$ are the indices of two data points in a high-dimensional space and $\delta_{ij}$ is the distance between them. In the context of names, we refer to \textit{dissimilarities} because some measures are not strictly distance metrics. For a given dissimilarity matrix \mbox{$\Delta=[\delta_{ij}]$}, a configuration matrix \mbox{$\mathbf{X}=[x_1,x_2,..,x_n]$}, where $x_i \in \mathbb{R}^p$ is constructed by minimizing 

\useshortskip

\begin{align}  \label{eq:1}
\sigma_{raw}(\mathbf{X}) &={\sum_{i,j=1}^n w_{ij}\Big(d_{ij}(\mathbf{X})-\delta_{ij}\Big)^2}.
\end{align}

The Euclidean distance between $i^{th}$ and $j^{th}$ points in the lower-dimensional space is given  by $d_{ij}$ and $w_{ij}$ denotes the possible weights from each pair of points. These non-negative weights indicates the importance of the residuals  \mbox{$d_{ij}(\mathbf{X})-\delta_{ij}$} of object pair $ij$. Weights are useful when we have input data with missing values. Since there is no restriction on any distance \textbf{X}, we can define fixed values of \mbox{$w_{ij} = 0$} if $\delta_{ij}$ is missing and \mbox{$w_{ij} = 1$} otherwise~\cite{MDS}.
 
One can use squared distances or a normalized raw stress criterion as well. We prefer the raw stress here since it is popular and theoretically justified~\cite{Bae:2010}.

We start with a matrix of dissimilarities between surnames in our dataset, then try to find the Euclidean embedding that minimizes the approximate distances in a least-squares sense. Usually, MDS has been used to find a visual representation or to discover clusters of high dimensional data in two or three dimensions. But we are using MDS to uncover whether a simple vector representation would be able to approximate the characteristics of a namespace and how accurate would that be in a lower dimension.

To that end, we need to determine an appropriate dimension~$p$, in which distances are maintained to some level of accuracy. We used two approaches to discover a reasonable lower dimension that fits our data.  First, we analyzed the rate of change of the \textit{stress} values with respect to $d$-dimensions. Lower stress values indicate better approximations. Second, we use a Shepard diagram analysis to determine the goodness-of-fit of LSMDS. A Shepard diagram is a popular way of assessing the goodness of fit of data reduction techniques such as MDS and t-SNE~\cite{vanDerMaaten2008}. It compares how far apart our data points are before and after transformation using a scatter plot. We present the results in Section 5.

\subsection{Multivariate Normal Analysis}

If we can approximate a namespace in a $p$ dimensional Euclidean space, this leads us to our second question, ``How do we simulate name-like vectors using the approximate namespace?" A simple approach is to determine the distribution of the projected lower-dimensional vectors and use it to simulate vectors with similar characteristics.

We start by checking the normality of the projected lower-dimensional vectors because it is generally a good starting point and the normal distribution makes simulation easy, requiring only two parameters: mean and covariance. Upon this basis, we started exploring the multivariate characteristic of the real namespace and the transformed Euclidean space.

The first step of the Multivariate Normal (MVN) analysis is to check the normality of the transformed vectors in the Euclidean space. The normal distribution (see Section 5) appears to be a reasonable model and so the next is to consider its characteristics. Thus we evaluate the multivariate normal characteristics of each variable of the transformed vectors in the Euclidean space. Histograms, box plots and \mbox{qunatile-qunatile} plots (Q-Q plots) are used to visually assess the characteristics of the distribution~\cite{Korkmaz2014}.

The observations suggest that the transformed vectors in the Euclidean space do not strictly follow a normal distribution. But in this study, we are looking for a simple model to generate the name-like vectors. If we go down the path searching for other distributions, then the simulation becomes more complex. Even the corrections to the normal distribution would be complicated. The trade-off between simplicity and accuracy leads to us using the normal distribution to simulate vectors in our model.

Name-like vector simulation is a two-step procedure. The first step is estimating the parameters (mean and variance) for the normal distribution that we would choose as our simulation model. Here, our sample is a Euclidean namespace of 5000 transformed name-like vectors. Since our data is multivariate, we calculate the sample covariance matrix ($\sigma$) \cite{JChatfied1980}. The second step is the simulation of name-like vectors using the estimated parameters of the normal distribution $N(\overline{x},\Sigma)$ where $\overline{x}=0$ . 

\noindent Once we simulate name-like vectors using the above two-step procedure, we explore the characteristics of the real namespace, the transformed namespace and a simulated namespace. Our main interest is on the distributions of the distances, as it is a key property of the namespace that we approximate in our simulation. We compare the three distributions of distances to evaluate the similarities between them and to see how well a normal distribution can reproduce the distances between vectors impersonating a real namespace.

The other characteristic we explore here is the errors of a real namespace and how these errors can be simulated in the transformed Euclidean space.

\subsection{Error Modelling}

One of the motivations for strong ER algorithms is that real-world datasets are low in quality. A namespace can contain typographical, OCR and phonetic errors or nicknames. Our goal is to include some of these errors in the simulated namespace.

Consequently, we come to the sub question: ``How do we model errors in a simulated namespace?" When simulating real name strings, the typical process would be to adopt an error generating model. In contrast, as we simulate name-like vectors, the aim is to generate errors by adding some noise to the simulated namespace. 

Errors can be thought of as variations of original names. Distances between a name and its variants are typically small compared to the distance between two distinct names. If we apply the same logic to name-like vectors, the distances between a name-like vector and that containing an error needs to be smaller. In our error model, we capture the variance of vectors and use it to simulate errors with small Euclidean distance away from the original. 

Our prime interest is edit distance errors such as insertions, deletions, transpositions and substitutions because they are the most common errors in real datasets~\cite{Christen2005}.

By looking at the variances of transformed vectors of errors and the original data vectors, we observed that the ratio between the covariance matrices is significantly low. This implies that we can add Gaussian noise to model some random errors in our simulated namespace.  It is a standard and a simple way of modelling random noise compared to complicated noise models. We generate this noise using a second normal distribution $N(0, \Sigma_e)$, different from the first one. We calculate the unbiased estimator of the covariance matrix ($\Sigma_e$) using the pooled covariance estimate $\hat{\Sigma_p}$.

The pooled variance is a method of estimating the variance of two or more populations by using the sample variances from these populations~\cite{Books}. The formula is given by
\begin{equation}
    \hat{\Sigma_p}= \frac{(n_1-1)\Sigma{_1} + (n_2-1)\Sigma{_2} + ... + (n_k-1)\Sigma{_k}} {n_1 + n_2+.. + n_k - k},
\end{equation}

\noindent where $ \hat{\Sigma{_p}}$ is the pooled covariance, $\Sigma{_1},\Sigma{_2},..,\Sigma{_k}$ are sample covariance and $(n_1-1),(n_2-1),..,(n_k-1)$ are the degrees of freedom. 

The expectation is to have a much smaller variance in the distribution of noise compared to the initially simulated distribution of name-like vectors.  Thus by adding some noise to simulated name-like vectors, we can represent errors in such a way that they preserve small distances to their original name-like vectors.

\subsubsection{Comparing Covariance Matrices}

As mentioned before the deviation of the noise should be significantly lower or otherwise, the noise might overcome the pattern within the original data. We compare the covariance of the two groups: transformed error vectors and the transformed name-like vectors using the ratios of covariance. The value of the ratio quantifies the two covariances. It tells us how big or small the covariance of transformed errors compared to the covariance of the transformed name-like vectors.

For $p$ variables, we have a symmetric $p \times p$ covariance matrix. Let $\Sigma_e$ and $\Sigma_s$ be the $p \times p$ covariance matrices of two groups. Then the ratio can be written as a product of one covariance matrix and the inverse of the other covariance matrix ($\Sigma_s^{-1/2}\Sigma_e \Sigma_s^{-1/2}$), and it tells us how to transform one to another. Hence this is a useful way to report the differences between the two \mbox{variance-covariance} structures. Relative eigenanalysis or relative principal component analysis (RPCA) determines the direction along which the ratio of variances between two groups is a maximum~\cite{Books}.

Since it is hard to compare covariance matrices in $p$ dimensions, we use RPCA on the matrix product described above to project them down to one or two dimensions. For instance, if we represent the matrix product as an ellipse (in two dimensions) the length of the axes of this ellipse equals the ratio of variances of the two groups in that direction. This direction is the first eigenvector of $\Sigma_s^{-1/2}\Sigma_e \Sigma_s^{-1/2}$,  also known as the first relative eigenvector of the $\Sigma_e$ with respect to $\Sigma_s$. 

The eigenvalues of $\Sigma_{s}^{-1/2} \Sigma_e \Sigma_{s}^{-1/2}$ are
called the relative eigenvalues of $\Sigma_e$ with respect to $\Sigma_s$. Thus the first relative eigenvalue $\gamma_1$, is equal to the maximal ratio of variances. Small $\gamma_1$ values indicate, even in the worst-case scenario, the errors have small variances. Thus we can assume, our error simulation model based on the Gaussian noise is approximately good. Our experiment results support this expectation.

In order to evaluate the potential benefits of the proposed solution, a set of experiments has been conducted on real datasets.

\section{Experimental Results}

This section presents the results of the ideas introduced in the previous section to real data, in particularly the 250,000 distinct surnames obtained from the Ancestry.com~\cite{7023436}. The final results do not depend on the individual observations in the random sample, because all of the random samples we considered gave us similar results. Hence the sample selection is flexible. In this paper, we have used randomly selected 5000 surnames for testing. The pairwise dissimilarities between surnames are calculated using a set of standard string differences.


\setlength{\belowcaptionskip}{-2ex}
\begin{figure}[ht!]
\centering
\def\big{\includegraphics[height=2.5in,width=3.55in]{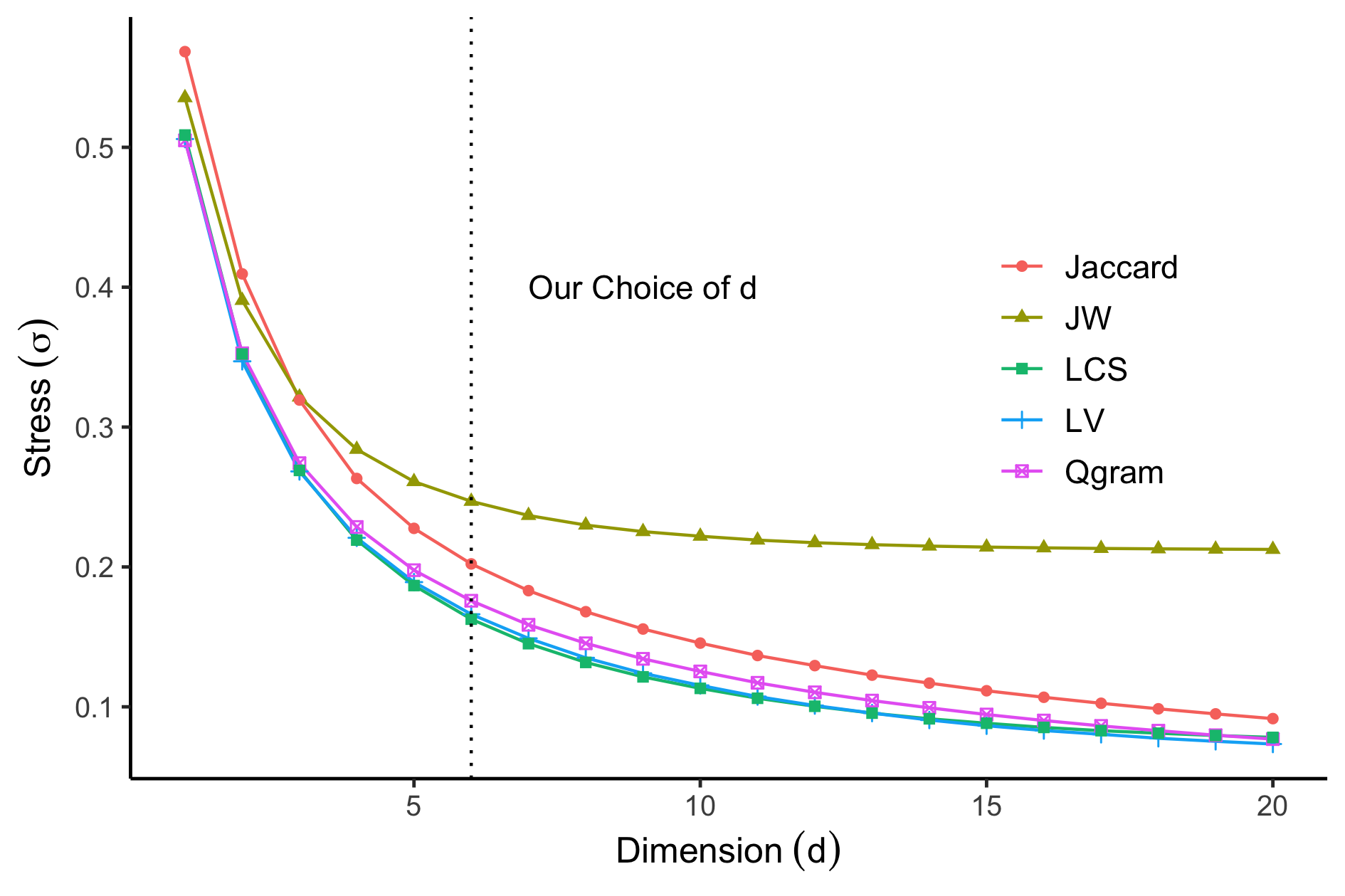}}
\def\little{\includegraphics[scale=0.11]{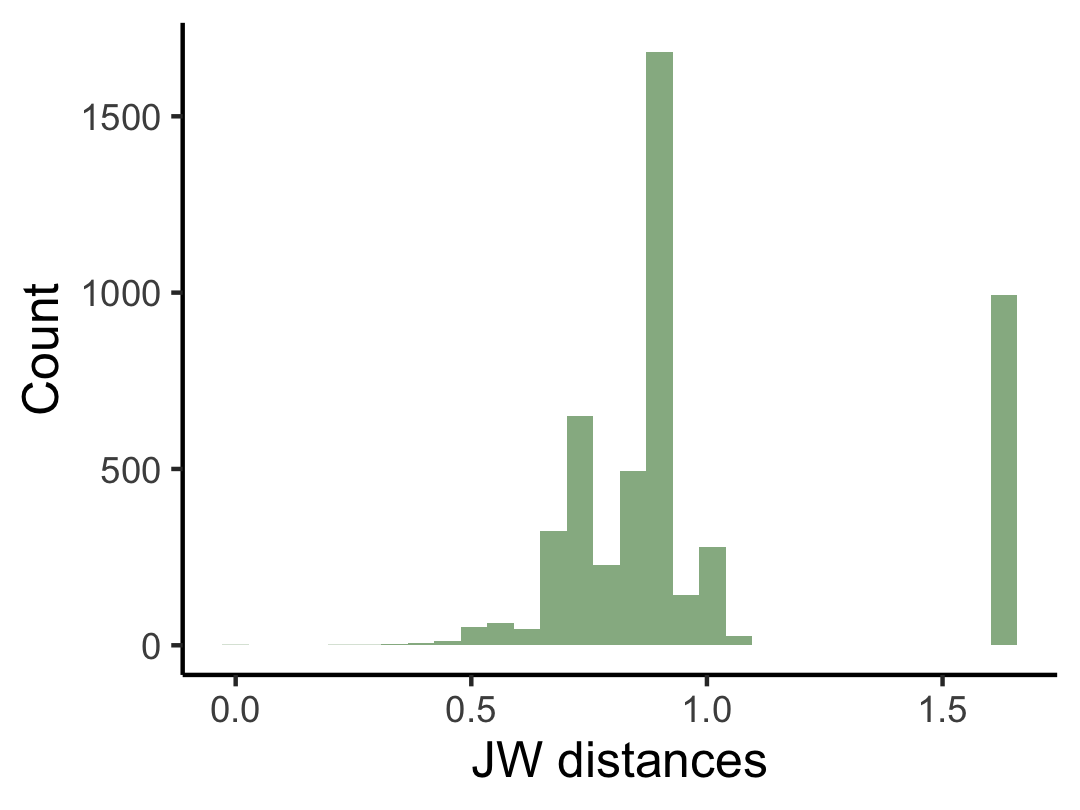}}
\topinset{\little}{\big}{-45pt}{50pt}
\caption{Values of the goodness of fit measure: \textit{\lq stress'} changes with the dimension for each of the five dissimilarity measures Jaro-Winkler (JW), Jaccard coefficient, Longest common subsequence (LCS), Levenshtein (LV) and Q-grams. The inset Figure shows the distribution of JW distances. Compared to the other distributions of distances, JW distances exhibits a spike where all the dissimilar surnames take one value.}
\label{fig:1}
\end{figure}

We use the STRINGDIST R package~\cite{Loo2014} to calculate the dissimilarities. Two widely used optimization algorithms implement LSMDS: the SMACOF~\cite{Jan2009} algorithm and the gradient descent algorithm~\cite{MIZE}. We compared both methods using \textit{normalized stress} and found no difference except SMACOF takes much longer. Thus we use \mbox{gradient-descent-based} LSMDS for further analysis.

The \textit{stress} is a measure of the goodness-of-fit of the LSMDS given the dimension. From now on, when we refer to \textit{stress}, it is the \textit{normalized raw stress} ($\sigma$). Lower \textit{stress} values indicate the LMDS approximation for the real data is good while higher \textit{stress} values indicate the approximation is poor.

\subsection{Stress analysis}

Initially, we considered five frequently used string dissimilarity measures~\cite{Cohen2003,Christen2006AIssues}: Levenshtein distances, Longest common subsequent (LCS), Jaccard dissimilarity, Q-grams and the Jaro-Winkler distances.

\autoref{fig:1} shows the rate of change in the \textit{stress} ($\sigma$) against dimension ($d$). The main observation for all the dissimilarity measures is that, when $d$ increases, $\sigma$ decreases rapidly at first and then more slowly. There is a drastic decrease in the $\sigma$ from 1-6 dimensions. Most $\sigma$ values continue to decrease but Jaro-Winkler (JW) distances are different. For most dissimilarities,  the stress tends towards a small, but non-zero asymptote, reflecting the non-Euclidean nature of the original data.

The JW distances have, however a much higher asymptote. Compared to the other distributions of distances, JW distances exhibits a spike (see inset) where all the dissimilar surnames take one value, which is very hard to represent in a Euclidean space. Hence we argue that the approximation of a namespace to a metric space using LSMDS works for all but the JW dissimilarities.

Ideally, we seek for an elbow in the \textit{dimension vs stress} plot, where there is an obvious point at which the best trade-off between stress and dimension can be found. In practice, however, such elbows are rarely obvious, so we aim to find a reasonably compact dimension. However, when we have non-zero \textit{stress}, we should keep in mind that the distances among vectors are imperfect, distorted, representations of the relationships given by our data.

We compared the results of the stress analysis for 10,000, 15,000 and 20,000 of test data sets to see the performance of the LMDS approximation. The Comparison shows that more data do not significantly increase the \textit{stress} values. Thus, for further testing, we will continue to use a fixed sample size of 5000 surnames. 

\subsection{Shepard diagram analysis}

A Shepard diagram~\cite{vanDerMaaten2008} compares the relationships between our data points before and after transformation. An accurate dimension reduction will produce a straight line that goes through the origin. However, in practice, Shepard diagrams rarely look straight due to information loss during data reduction.

\setlength{\belowcaptionskip}{-1ex}
\begin{figure}[ht!] 
\centering
\includegraphics [height=3in, width=3.5in]{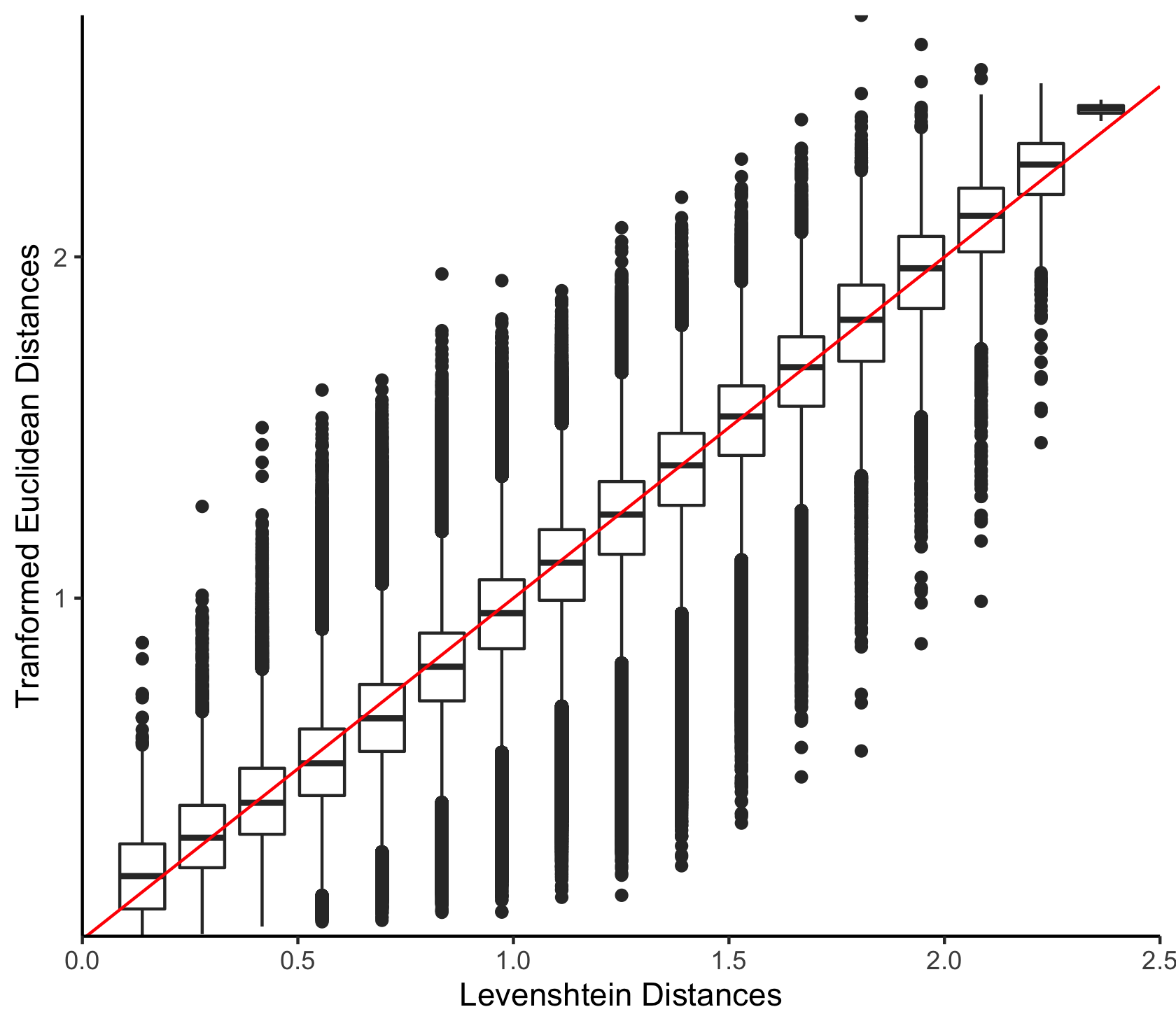}
\caption{The Shepard diagram \cite{vanDerMaaten2008} of the Levenshtein distances (LV) vs Euclidean distances in a 6-\textit{d} space. There is a strong linear relationship between the two distance values as desired.}
\label{fig:2}
\end{figure}

\autoref{fig:2} shows a Shepard diagram obtained for the Euclidean distances against the LV distances. The Euclidean distances are measured in a 6-\textit{d} space.

We use a boxplot to show the variation in the data around the straight line, which shows a strong linear relationship between two distances distributions. We saw that the \textit{stress} is not zero from the previous subsection, so there are some errors in the approximation. 

We compared Shepard diagrams obtained for 8-\textit{d}, 10-\textit{d}, 15-\textit{d} and 20-\textit{d} for the same dataset to see the performance of the approximation as the dimension of the target space increases. The comparison confirms that more dimensions do not significantly increase the linearity of the LSMDS approximation. The variations around the line decrease, but the line does not fit better. Considering the trade-off between the simplicity of our simulation model and the accuracy of the reproduced distances, we select the 6-\textit{d} Euclidean space as the basis for our simulation model.

We prefer LV distances for further analysis because it is popular among string dissimilarity metrics and performs well with the two goodness-of-fit methods. 

However, we tested the others with similar results except for the JW distances, as noted earlier. The approximation fails to recover a reasonable linear relationship between JW distances and the Euclidean distances due to the higher \textit{stress} values.  

\subsection{Multivariate Normal Analysis}

We assess the multivariate normality of the transformed vectors in the Euclidean space using multivariate normality tests, multivariate plots, multivariate outlier detection, univariate normality tests and univariate plots~\cite{Korkmaz2014}.

Initially, we perform multivariate skewness and kurtosis tests for the transformed vectors in the Euclidean space and combine test results for multivariate normality. Both tests fail to indicate multivariate normality for the transformed vectors.

\setlength{\belowcaptionskip}{-2ex}
\begin{figure}[ht!] 
\centering
\includegraphics [height=2.5in, width=3.5in]{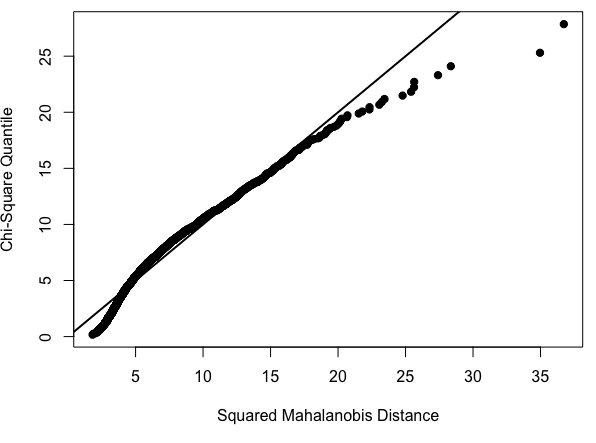}
\caption{
Shows the chi-square Q-Q plot for the distribution of the all 6-dimensional vectors. There are some deviations from the straight line indicating the possible departures from a multivariate normal distribution.}
\label{fig:3}
\end{figure}

To understand why the tests fail, we use a Q-Q plot as shown in \autoref{fig:3}. If the transformed vectors follow a multivariate normal distribution, the points in the Q-Q plot should lie approximately on the $y = x$ line. As the \autoref{fig:3} indicates, there appear a few, large squared Mahalanobis distances (outliers) suggesting possible departures from a multivariate normal distribution in the upper tail in particular. However, the body of the distribution lies largely on the $y = x$ line. This suggests that normality is a reasonable model for most of the data.

\setlength{\belowcaptionskip}{-2ex}
\begin{figure*}[ht!] 
\includegraphics [height=4in, width=6in]{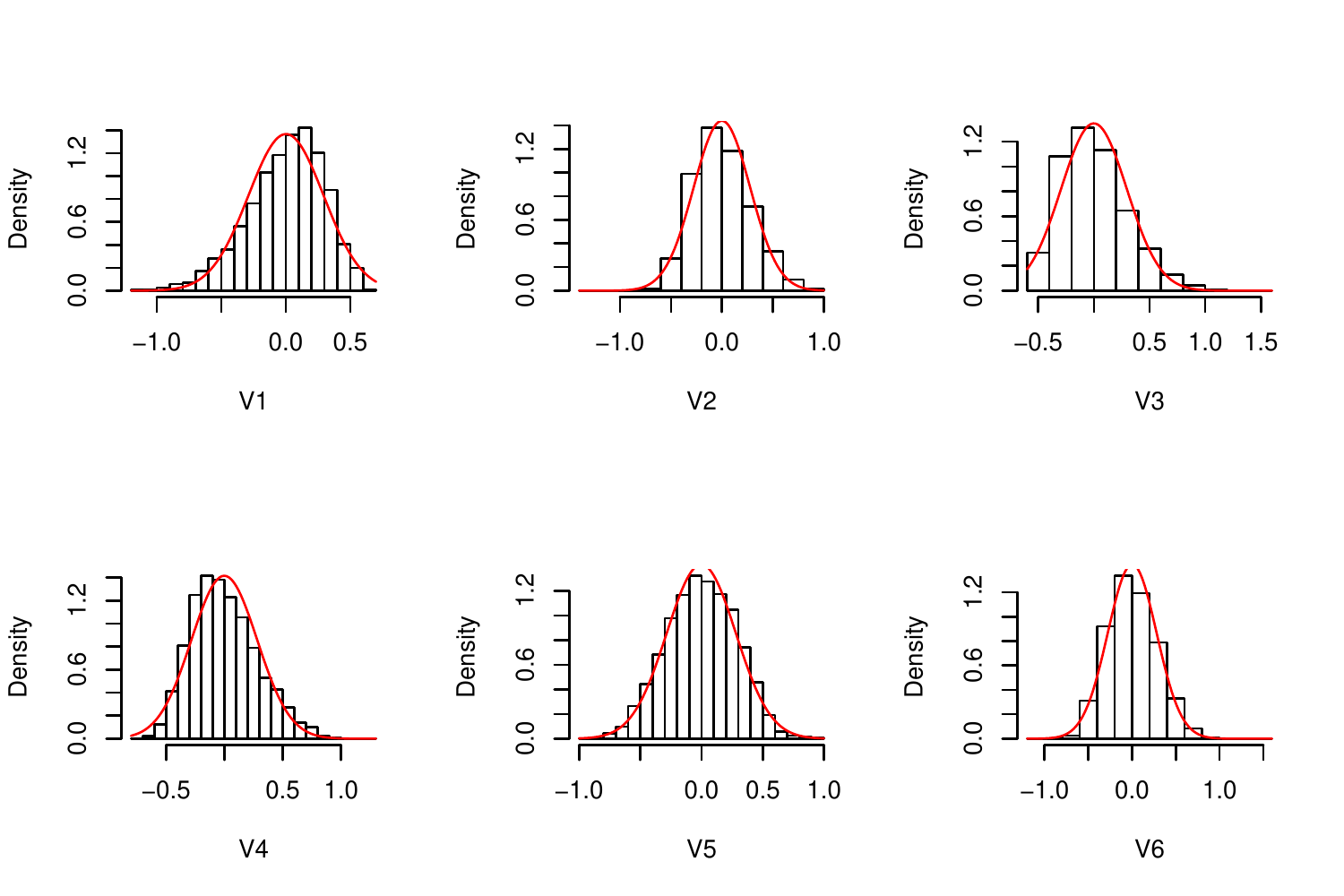}
\caption{Histograms of the transformed coordinate vectors. The variables V5 and V6 of the coordinates shows an approximately normal distribution while the others have slightly skewed distributions.}
\label{fig:4}
\end{figure*}

To diagnose the reason for the deviation from MVN, one can perform univariate tests. Furthermore, checking univariate plots can also be very useful. In the univariate setting, strict normality tests such as Kolmogorov-Smirnov (K-S) and Shapiro-Wilk are widely used in practice~\cite{Korkmaz2014}. But these tests reject the normality assumptions for each of the variable in the transformed vectors.

We also used histograms and Q-Q plots to assess the univariate normality. \autoref{fig:4} shows the histograms of the transformed coordinate vectors considering each variable. The relevant Q-Q plots are not included due to space limitation. However, the Q-Q plots and the histograms suggest that the variables 5 and 6 are quite well approximated by normal distributions while the other variables have slightly skewed distributions, but the body is still moderately normal.

The problems with multivariate normality seem to be relatively minor. Given the simplicity of generating multivariate normal variables compared to almost any other multivariate distribution, and to the relatively small deviation from this in the data, we choose to use the normal distribution in our synthesis method.

\subsubsection{Estimation and Simulation}

Given that we approximate the data as normal we need to estimate parameters. In the first step, we estimate the covariance matrix ($\Sigma$) from the transformed vectors in the 6-\textit{d} space. The resulted covariance matrix has values very near to zero on off-diagonals. We test for independence between each variables using correlation test and Hoeffding's D statistics~\cite{Hoeffding1994}. Both approaches verify that we can assume the independence between the 6-\textit{d} variables.

We simulate the name-like vectors using the estimated value of the covariance matrix ($\Sigma$) for the normal distribution of $N_p(0, \Sigma)$ where $p$ is the dimension. Then a simulated surname (a name-like vector) can be represent using $v_1,v_2, v_3, ..., v_p$.

In the next steps, we look at the characteristics of the simulated namespace compared to a real namespace, \textit{e.g.,}~distances and errors.

\subsubsection{Comparing distances distributions}
\setlength{\belowcaptionskip}{-2ex}
\begin{figure}[ht!] 
\centering
\includegraphics[height=2.4in, width=2.4in]{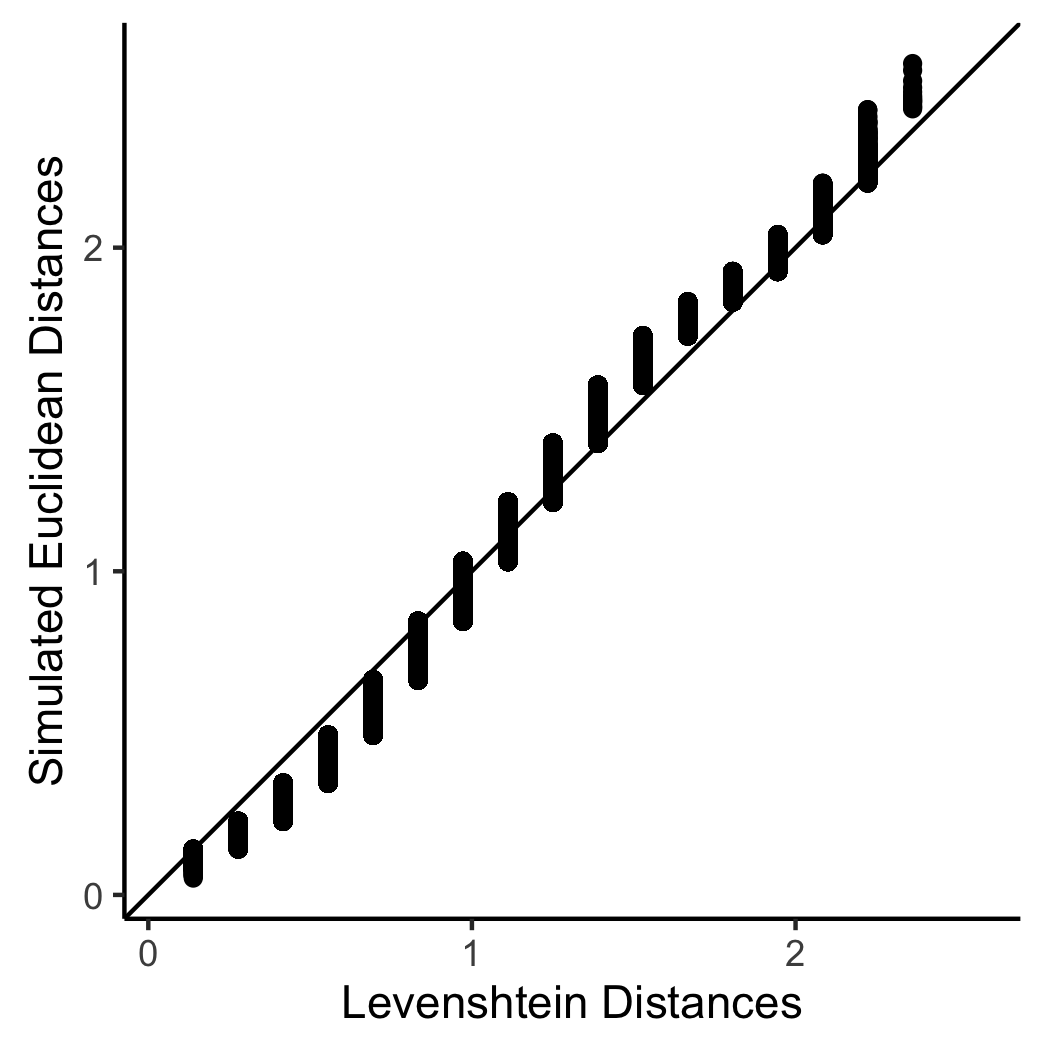}
\caption{Q-Q plot compares the distributions of LV distances with the Euclidean distances of simulated vectors. Points along the straight line indicate a similarity agreement between the two distributions.}

\label{fig:5}
\end{figure}

\begin{figure*}[ht!] 
    \includegraphics [height=4in, width=7in]{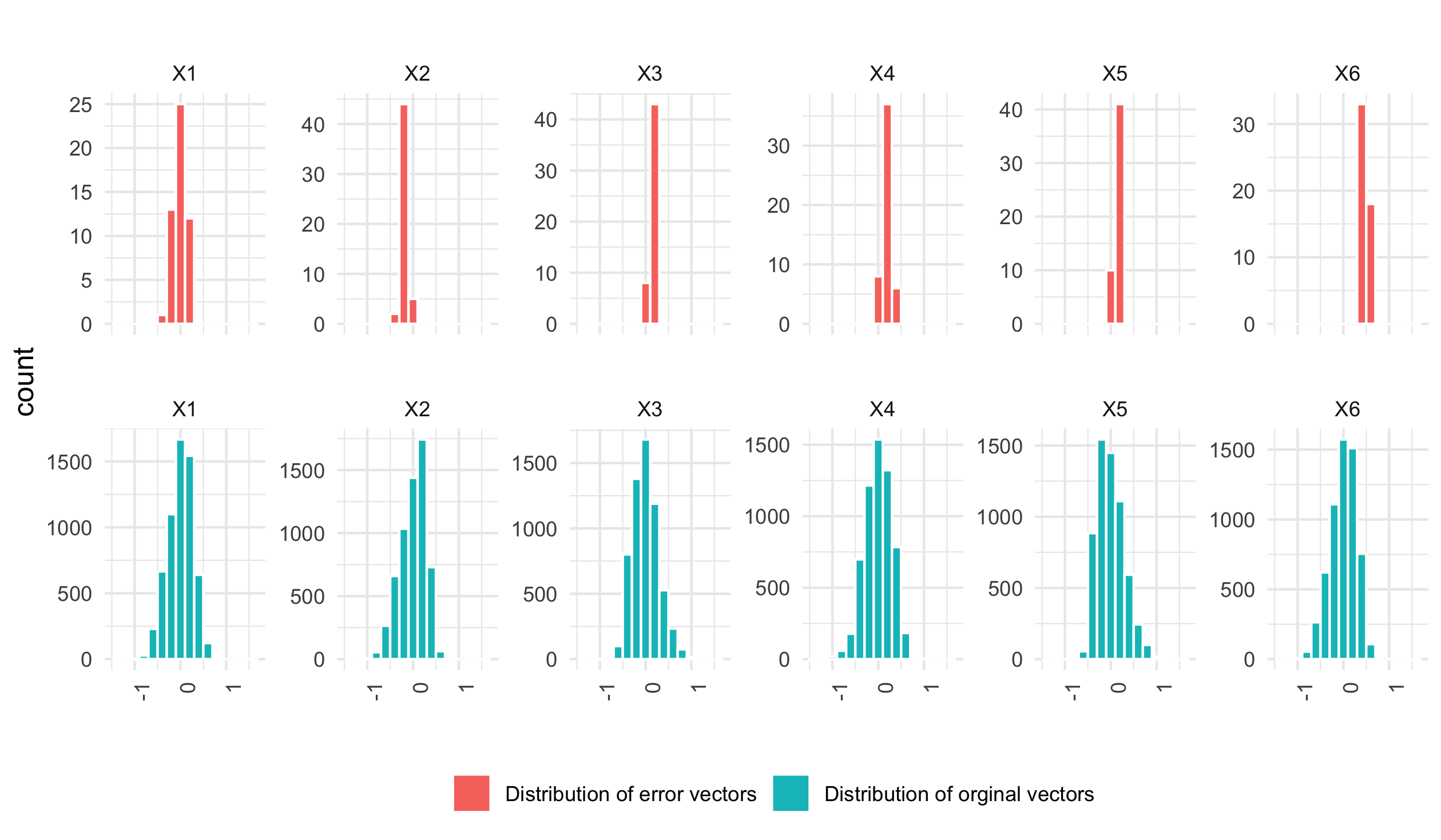}
    \caption{ The top row shows the spread of the coordinates of each errored name around the correct surname for each coordinate in 6-\textit{d} space. The bottom row shows the spread of the coordinates of each prediction variable in the whole sample, including all the surnames and their errors. The comparison shows that the spread of errors is much smaller compared to the spread of surnames and their errors.}
    \label{fig:6}
\end{figure*}    

The general aim of the proposed simulation model is to generate name-like vectors that have some key properties of a real namespace. The distances or the dissimilarities between surnames are of the prime interest here. One further test is to now compare the distribution of distances created by the simulation model to those of the original data.

\autoref{fig:5} shows the Q-Q plot that compares distributions of LV distances between surnames and  Euclidean distances between simulated name-like vectors. The Q-Q plot exhibits a reasonable similarity between the distributions even if they are not exactly falling along the $y=x$ line. We can see the discrete nature of the LV distances and the continuous Euclidean distances as they give the points along small lines.

\noindent
There is some distortion created by the transformation process, but apart from this, the simulated vectors behave very similarly. The conclusion is to assume that the distributions are quite similar.

\subsection{Error simulation}

By extending our primary data analysis, we explore the nature of the error variance in a real namespace to simulate it in the Euclidean space. First, we look at the variance of errors once they are transformed into a 6-dimensional space. Then we compare the variance of errors and name-like numbers in the Euclidean space using the ratios of covariances in the two types of vectors.

We select a random sample of 20 surnames from the initial dataset of 5000 surnames and generate 50 variations of edit distance errors for each of them. The errors are generated using FEBRL~\cite{Christen2008}. Then we apply LSMDS on the 5000 distinct surnames + 1000 errors. We use these transformed vectors to explain the variance of the errors in a Euclidean space.


We then examine the variance of errored names relative to their correct version. We contrast this variation with the Euclidean distances in the whole sample. \autoref{fig:6} illustrates a comparison between the histograms of the whole sample (surnames + errors) with 50 variations of errors belongs to one surname. The histograms show a small spread of the vectors that represent errors compared to the spread of the vectors in the whole sample.

The result implies that we can add appropriate noise by adding a small relative variance to the original set of name-like numbers to represent errors in the Euclidean space. 

We also investigate the variance-covariance matrices correspond to each of the erroneous surnames and their variations. These values were much smaller compared to the variance-covariance matrix of the whole sample which indicates that these small edit distance errors are closer to their real surname values than other names. Thus we can generate errors by adding some noise from a second normal distribution $N(0, \Sigma_e)$ to the simulated namespace. The values of the $\Sigma_e$ can be estimated using the pooled covariance, which is the weighted average of all the variance-covariance matrices correspond to each of the erroneous surnames and their variations of the chosen random sample.

The maximal ratio of the covariance matrices $\Sigma_e$ and $\Sigma_s$ takes a value closer to 0.1. To remind the readers,  $\Sigma_e$ represent the covariance matrix of transformed error vectors and $\Sigma_s$ represent the covariance matrix of the transformed name-like vectors. The maximal ratio measure the maximum error variation with respect to the variation of transformed name-like vectors. Since we get a small value for the maximal ratio, we conclude that even in the worst case, the errors have small variance compared to the name-like vectors in a Euclidean space. Thus, our model can simulate simple edit distance errors by adding some Gaussian noise to an existing Euclidean metric space. 

\subsection{Comparing the computational time}

In this section, we discuss the practical use of our simulator for developing big data ER algorithms. One of the dominant issues in ER algorithms is scalability. Since the number of pairwise comparisons between records is quadratic, it is prohibitive for large datasets. 

It is challenging when the majority of the records contains strings. But with the name-like numbers, even for a full set of pairwise comparisons, the computational time is much lower.

\setlength{\belowcaptionskip}{-1ex}
\begin{figure}[ht!] 
\centering
\includegraphics[height=2in, width=3.2in]{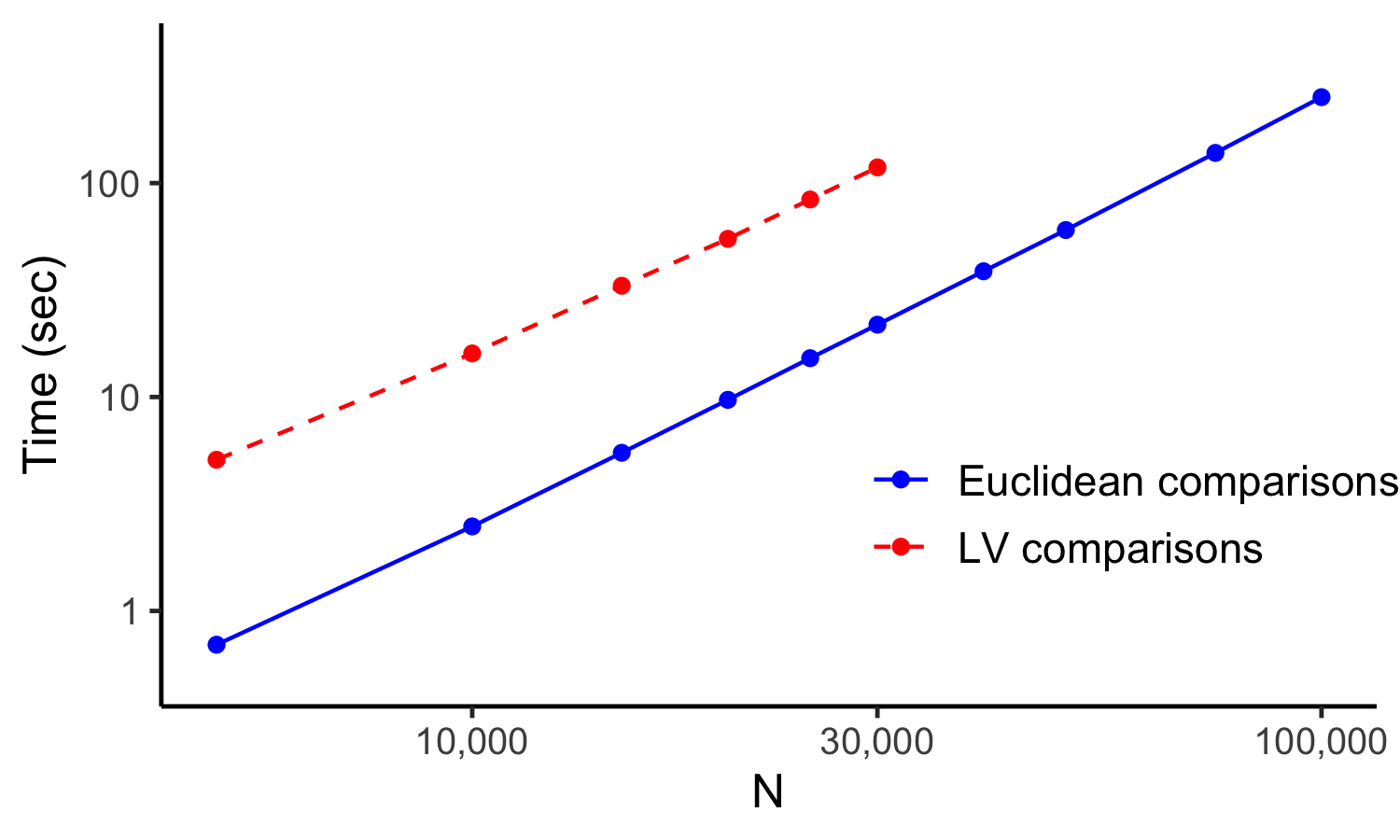}
\caption{Compares the computational time taken for calculating pairwise distances between name strings and name-like numbers. N is the number of data points in each instance. Calculating the Euclidean distances is much faster than calculating LV distances.}
\label{fig:7}
\end{figure}

We compare the computational time of the pairwise comparisons between a dataset contains real surnames with a simulated set of name-like vectors. \autoref{fig:7} shows a comparison between the two approaches on log-log axis. The surname comparisons are based on the LV distances, while name-like vector comparisons are based on Euclidean distances. 

The time grows quadratically for both approaches, but the difference is significant. Pairwise comparisons between name-like vectors are an order of magnitude faster compared to the surname comparisons. Thus our simulation data is better for experiments with large datasets in big data ER.

\section{Discussion and Future Work}

Using our simulation model, we can synthesize very large datasets of name-like vectors. Our simulation model is a simple, inexpensive and fast approach to simulate names, but the model has its limitations.

Our simulator does not generate complete records. Surnames are just a one piece of information required for linking data. To get the full benefit of the proposed simulator, we still need to model other linking variables such as addresses, gender. As a result, the dimensions of the vectors need to extend accordingly to facilitate complete records. However, numerical or categorical data such as age or gender are easier to simulate.

Real name spaces also have other interesting characteristics. For instance, phonetic errors and variations of the same name or nicknames. We are not addressing these issues here.

Furthermore, surnames (or names) follow a Zipf distribution~\cite{zip}. In a Zipf distribution, few high-frequency names account for most of the population with many low-frequency names. Our model has not captured the frequencies of names when simulating name-like vectors. But this is easier to address by repeated sampling.

A key advantage is that the approach is not intrinsically limited to a particular culture or language. So should be more widely applicable than resampling techniques. Also the nature of the data allows an arbitrarily large number of synthetic data points. For very very large datasets, one can also increase the dimension of the space to reduce the density which it is populated. The volume of such a space increases exponentially with dimension.

Towards the developing of ER techniques for big data, our focus is on the global or the holistic record matching. We are interested in extending this work to solve the global matching problem for large datasets. ER has been studied for years, but the focus is on pairwise comparisons. For large data sets, this is where the bottleneck occurs as the naive comparisons between records become quadratic. Avoiding the comparisons of unnecessary detail in the pairwise comparisons can make the ER process more scalable. Simulation of the many details of identification keys is not required when we are considering the global matching problem.  

\section{Conclusion}
Our goal here was to generate synthetic data to develop and test ER algorithm appropriate for big data. We proposed a simple, inexpensive and fast simulation model that can generate name-like vectors including simple errors. In this paper, we discuss how to simulate simple vectors in a space that approximates the properties of names as one step towards being able to generate large simulated datasets for large-scale testing of global matching techniques.

\ifCLASSOPTIONcaptionsoff
  \newpage
\fi

\bibliographystyle{IEEEtran}
\bibliography{IEEEabrv,ref}


\end{document}